\documentclass[onecolumn,aps,showpacs,]{revtex4}
\usepackage{graphicx}
\input{epsf}

\begin{document}

\title{High Bias Voltage Effect on Spin-Dependent Conductivity and Shot Noise in Carbon-doped Fe(001)/MgO(001)/Fe(001) Magnetic Tunnel Junctions}

\author{R. Guerrero, D. Herranz, F.G. Aliev}
\affiliation{Departamento de Fisica de la Materia Condensado, C-III, Universidad Autonoma de Madrid, 28049, Madrid, Spain}
\author{F. Greullet, C. Tiusan, M. Hehn, F. Montaigne}
\affiliation{Laboratoire de Physique des Mat\'eriaux, UMR CNRS
7556, Nancy Universit\'e, Bd. des Aiguillettes, B.P. 239, 54506 Vand{\oe}uvre-l\`es-Nancy Cedex, France}

\date{\today}

\begin{abstract}

Low temperature (10K) high voltage bias dynamic conductivity (up to
2.7V) and shot noise (up to 1V) were studied in epitaxial
Fe(100)/Fe-C/MgO(100)/Fe(100) magnetic tunnel junctions, as a
function of the magnetic state. The junctions show large tunnel
magnetoresistance (185\% at 300K and 330\% at 4K). Multiple sign
inversion of the magnetoresistance is observed for bias polarity
when the electrons scan the electronic structure of the bottom Fe-C
interface. The shot-noise shows a
Poissonian character. This demonstrates a pure spin dependent
direct tunneling mechanism and validates the high
structural quality of the MgO barrier.

\end{abstract}

\maketitle

Magnetic tunnel junctions (MTJs)\cite{Moodera95,Miyazaki95} are nowadays one of the most active areas of material science and spintronics. Recent theoretical predictions\cite{Butler01,Mathon01} and experimental demonstration\cite{Bowen01,FVincent03,Parkin04,Yuasa04,Tiusan06} of coherent spin-dependent tunneling in single crystal Fe(100)/MgO(100)/Fe(100) MTJs revolutionized this area providing new ways to create devices with room temperature Tunneling Magneto-Resistance (RT-TMR) exceeding 100\%. The large TMR at low bias is mostly due to fully spin polarized $\Delta_1$ bulk electron states in Fe(001), reflected for antiparrallel ferromagnetic electrodes configuration (AP) or well transmitted for the parallel (P) state\cite{Butler01,Mathon01}. However, the tunneling mechanism gets more complex when taking into account the electronic structure of the interfaces\cite{TiusanPRL04} and when biasing the junction. Therefore, for finite bias polarities the antiparallel conductance may exceed the parallel one, resulting in TMR suppression\cite{Yuasa04} or its sign reversal\cite{TiusanPRL04}. By engineering the chemical and electronic structure of the Fe/MgO interface, the voltage variation of the TMR in amplitude and sign can be skilfully manipulated. It has been recently demonstrated that the carbon-doping of the bottom Fe/MgO interface leads to strongly asymmetric TMR vs bias, providing a root for creation of high-output voltage device applications\cite{Tiusan06}.

Our Letter presents a first study of dynamical conductance and TMR
in a large bias window, up to 2.7 V, for
Fe(100)/Fe-C/MgO(100)/Fe(100) MTJs. The shot noise analysis in
different magnetization configuration is performed at voltages up to
1V. The experiments are done at room temperature (300K) and
low temperature (4K-10K). The measured TMR ratio increases from
185\% at 300K to 330\% at 4K, mostly due to the strong temperature
variation of the tunnel conductivity in the antiparallel (AP) state.
Moreover, our tunneling spectroscopy experiments show a clear
maximum in the AP conductivity for a finite bias and a multiple TMR
sign inversion. These experiments demonstrate the role of the
minority spin Fe interface resonance state (IRS) to the tunneling.
Furthermore, in both parallel (P) and antiparallel magnetization
configuration, the shot noise measurements demonstrate an
uncorrelated direct tunneling mechanism across the MgO barrier. The
shot noise analysis and the large breakdown voltage of the junctions
(up to 3V) demonstrates the high quality of our MgO barriers
(i.e. absence of defects such as oxygen vacancies).

Our epitaxial Fe(45nm)/ MgO(3nm)/ Fe(10nm)/ Co(20nm)/ Pd(10nm)/
Au(10nm) samples were grown by molecular beam epitaxy (MBE) on
MgO(100) substrates under UHV condition ($4*10^{-11}$ mbar base
pressure). Different coercive field of the MTJ electrodes is
obtained by hardening the top bcc Fe electrode with hcp Co
epitaxially grown with in-plane c axis. Prior to deposition, the
substrate is annealed at $600^\circ$, then the layers are grown at
room temperature. For flattening, the Fe electrodes are annealed to
$450^\circ$ (bottom Fe) and $380^\circ$ (top Fe). Following the
growth procedure\cite{Tiusan07}, two different samples can be grown:
samples with clean Fe/MgO bottom interfaces and samples with carbon
doping at bottom Fe/MgO interface (Fe/Fe-C/MgO). The Reflection
High-Energy Electron Diffraction (RHEED) analysis performed on each
layer of the MTJ stack allows a direct control of the epitaxial
growth and the high crystalline quality of the epitaxial layers.
Within the pseudomorphical growth regime, the two-dimensional
layer-by-layer growth of the MgO barrier is monitored by the
intensity oscillations in the RHEED patterns. Compared to clean
samples, in the samples with carbon the bottom Fe(001) electrode
presents a $c(2\times 2)$ surface reconstruction, as shown by the
RHEED picture (Fig. 1). After the growth of the multilayer stack,
MTJs with micrometric lateral size have been patterned using
standard optical lithography/ ion etching process. All the MTJs studied here contain
carbon doped Fe/MgO interface. They have shown a large voltage stability,
up to 3 Volts.

Dynamic conductance $G(V)$ and shot-noise bias dependence have been
studied using four-probe method with a set-up allowing to vary the
temperature between 2 and 300K, equipped with preamplifiers situated
on top of the cryostat. Two different techniques were employed to
measure dynamic conductance in P or AP states, providing nearly
identical results. The first one uses a current-voltage ($I-V$)
converter and a voltage amplifier. In this case, the MTJ is biased
using a constant \textit{DC} voltage with superimposed low amplitude
sinusoidal wave ($V_{AC}<20mV$). The voltage drop on the junctions
and the current were obtained by using an analogue-digital converter
(ADC) and a lock-in amplifier, providing the dynamic conductance.
The second technique, mainly employed at high bias, uses square
current wave of current superimposed on \textit{DC} current. Shot
noise measurements were done using a cross-correlation technique.
More details of experimental setup were published
elsewhere\cite{Guerrero05,Guerrero06}.

At 300K, the Fe/Fe-C/MgO(3nm)/Fe/Co MTJs show $R\times A$ product values (RT) ranging from 0.42 to 0.48 M$\Omega$.$\mu$$m^2$. The inset in the top panel of Fig. 2 shows typical TMR curves measured at 10mV either at 300K and at 4K. The large TMR ratio of 185\% at 300K indicates the high quality of the MTJs. Interestingly, the low temperature TMR ($\sim$ 330\%) notably exceed previously reported (250\%) maximum values of zero-bias TMR in epitaxial Fe(100)/MgO/Fe MTJs with 'undopped' Fe/MgO interfaces\cite{Yuasa04}.
The temperature variation of the TMR is understood from the dynamic conductivity experiments $G=dI/dV$ shown in Fig. 2(a) which
plots $G(V)$ at 300 and 10K within a voltage range of 0.8V. Firstly, asymmetric G(V) characteristics in positive and negative voltage demonstrate different electronic structure of the top and bottom electrodes and Fe/MgO interfaces\cite{Tiusan06}. Secondly, we remark significantly different temperature variation of conductivity in P and AP magnetization configurations. In the AP configuration (Fig. 2(b) bottom panel), we remark almost no temperature dependent shape variation, except the enhancement of low bias anomaly at 10K. However, we notice a strong reduction of $G_{AP}(V)$ by 50\% at low temperature. On the other hand, a net temperature dependent shape variation between 300K and 10K (Fig. 2(a) top panel) is clearly seen for $G_P(V)$.
Interestingly, the zero bias $G_P$ is mostly constant with temperature (only 2\% variation). Additional local minima appear at 10K for both positive and negative finite bias voltage. At low temperature, all studied MTJs reveal novel P-state low-bias conductance oscillations with about 4 minima (top panel of Fig. 2). We note that low-bias conductivity minima in the P state have been already observed in carbon free samples even at 300K. However, we always measured only two local conductance minima\cite{Tiusan07}. These minima were explained by the $\Delta_5$ majority electron contribution to the total conductivity at low voltage ($<0.3eV$ which is the top of the majority $\Delta_5$ band (Fig. 2(b)). The origin of low temperature $G_P(V)$ minima observed in Fe/Fe-C/MgO/Fe MTJs opens interesting theoretical perspectives. These calculations should investigate in detail the effect of the realistic electronic structure of the Fe/Fe-C/MgO (i.e. effects of Fe-C bounding) on the tunneling, in the low bias regime.

Figure 3(a) presents high bias conductance for voltages up to 2.7V,
measured at 10K. The influence of Joule heating (few K) on the
$I-Vs$ is neglected due to the rather weak observed low temperature
temperature dependence of both $G_P$ and $G_{AP}$. Interestingly,
while $G_{P}(V)$ is rather symmetric, in negative voltage when the
electrons tunnel into the bottom Fe-C/MgO electrode, the $G_{AP}(V)$
shows a strong asymmetric local maximum superimposed on roughly
parabolic background. This 'local' resonant increase of the $G_{AP}$
($G_{AP}>G_P$) in a narrow\cite{width} energy window will lead to
the lower voltage sign reversal of the TMR (Fig. 3(a), Bottom
panel).Similar to Scanning Tunneling Spectroscopy
Experiments\cite{Stroscio}, and as we already previously
shown\cite{TiusanPRL04}, the resonant enhancement of $G_{AP}$ is
attributed to the contribution to the tunneling of the Fe minority
interfacial resonance (IRS). However, we only observed this
phenomena in carbon free Fe/MgO/Fe samples with thinner MgO barrier,
where the Fe IRS still significantly contribute to the
tunneling\cite{IRS}. In the samples studied here, having carbon at
the Fe-C/MgO interface, an important effect of the $G_{AP}$ resonant
activation by IRS is observed even for 3nm thick MgO barriers.
To elucidate this interesting property, theoretical
investigation of two effects is in progress: (i) the effect of
Fe-C-MgO bounding on the minority spin Fe(001) IRS (i.e. shift in
energy, dispersion in k); (ii) the carbon induced periodical
perturbation of the potential at the bottom Fe/MgO interface (i.e.
$c(2\times 2)$ reconstruction, Fig. 1) induces scattering events
which change k-vector. This has direct consequences on the total
conductivity.

In positive bias, when electrons are injected toward the top electrode, the low bias TMR changes the sign above 1.5V. This is determined by the $G_{AP}$ strong enhancement when, in the AP configuration the injected $\Delta_1$ electrons from the bottom Fe electrode arrive as hot electrons in the top electrode and find an equivalent symmetry in the minority band. In negative voltage, when electrons tunnel into the bottom Fe-C/MgO electrode, similar contribution of the minority $\Delta_1$ symmetry to the conductivity is expected. However, the TMR second sign reversal seems to appear at much higher voltages, above 2.5V (Fig. 3(a), bottom panel). One possible reason would be the reduction of the hot electrons thermalization length in the bottom electrode. The effect of the IRS at Fe-C/MgO interface on this phenomena requires further theoretical investigation.

Figure 3(b) presents shot noise measurements carried out at T = 10K
on Fe/Fe-C/MgO/Fe MTJs, with bias direction corresponding to
injection of electrons from the top to the bottom (carbon doped)
Fe/MgO interface. For comparison, the solid curves shows the
'theoretical' expectation for the shot noise, for electron tunneling
having Poissonian character: $S_V = 2e<I>/{G}^2$  with $G$ the
dynamic conductivity (Fig. 3(a)) and $I$ the applied current. Within
the error bars, showing dispersion of the shot noise 'white'
spectrum in the kHz range, the experimental data clearly indicate
the absence of electron correlations and/or sequential tunneling
phenomena. This proves that both P and AP spin dependent
conductance's and the shot noise are due to direct tunneling between
electron bands, as expected for the coherent
tunneling\cite{Blanter00}. The absence of resonant assisted
tunneling in the shot noise demonstrates the high quality of our
epitaxial MgO barriers (i.e. the absence of oxygen vacancies). This
high quality is furthermore confirmed by the large breakdown voltage
of the MTJs (up to 3V).

The authors thank G. Lengaigne for technological process of MTJs and
R.Villar for discussions and continious support.

\section*{FIGURE CAPTIONS}

{\bf FIG.1}\\
RHEED patterns of the Fe bottom layer for (a) carbon free Fe and (b) Fe/Fe-C along the [110] crystallographic direction.
Additional pattern for Fe/Fe-C surface demonstrate the $c(2\times 2)$ reconstruction related to carbon.

{\bf FIG.2}\\
Color online: (a) Dynamic conductivities in P (top panel) and AP (bottom panel) magnetization states at 300K (open circles) and 10K (full circles). Top panel inset: TMR curves at 300K (red open circles) and 4K (black full circles). (b) Bulk band structure diagram of bcc Fe.

{\bf FIG.3}\\
Color online: (a) Dynamic conductivities at 10K (top panel) and
related TMR(V)(bottom panel). (b) Shot noise measurements in P and
AP states measured at 10K in bias when the electrons are injected
from the top toward the bottom MTJ electrode (negative voltage in Fig. 2 and 3.a).

\newpage

\begin{figure}[htbp]
\begin{center}
\includegraphics[width=7.5cm]{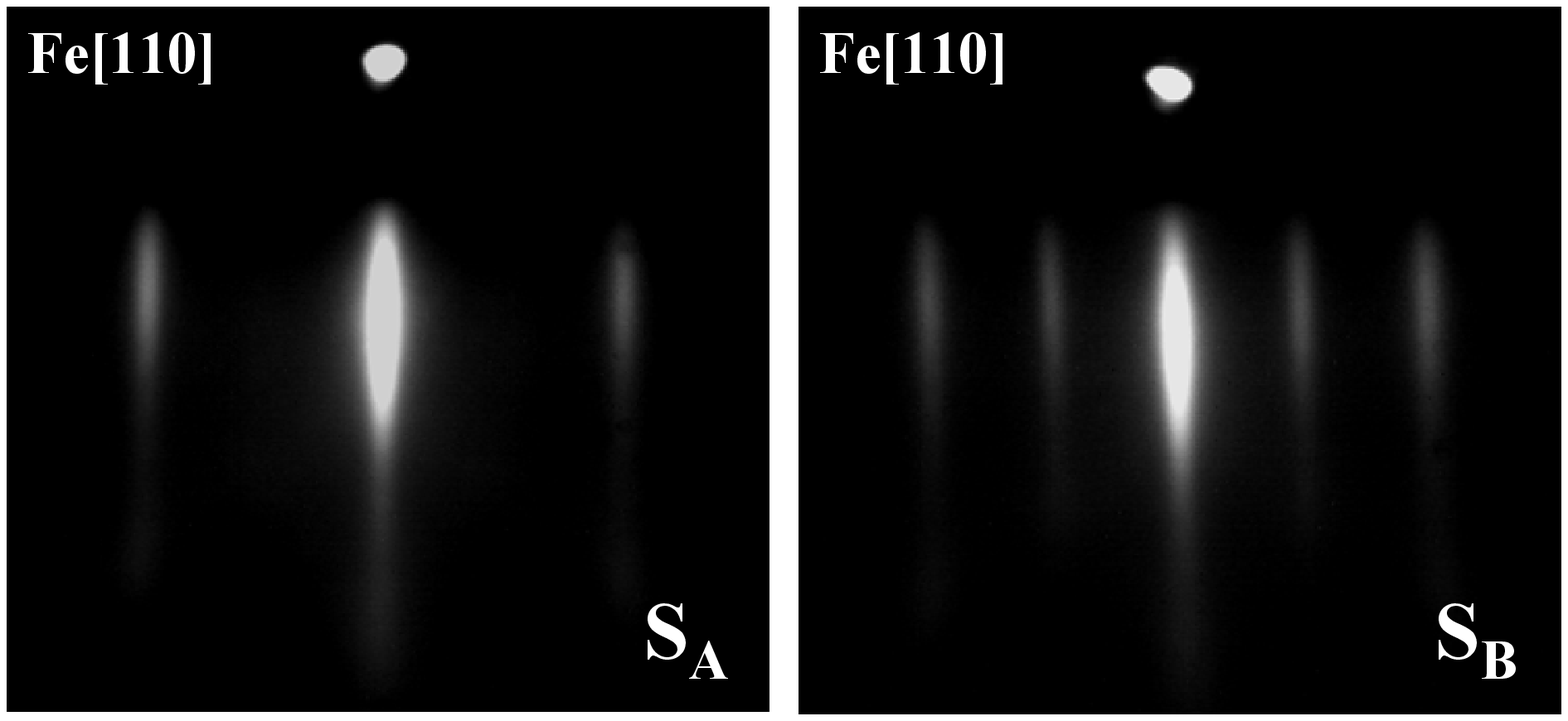}
\caption{}
\label{Fig1}
\end{center}
\end{figure}

\begin{figure}[htbp]
\begin{center}
\includegraphics[width=7.5cm]{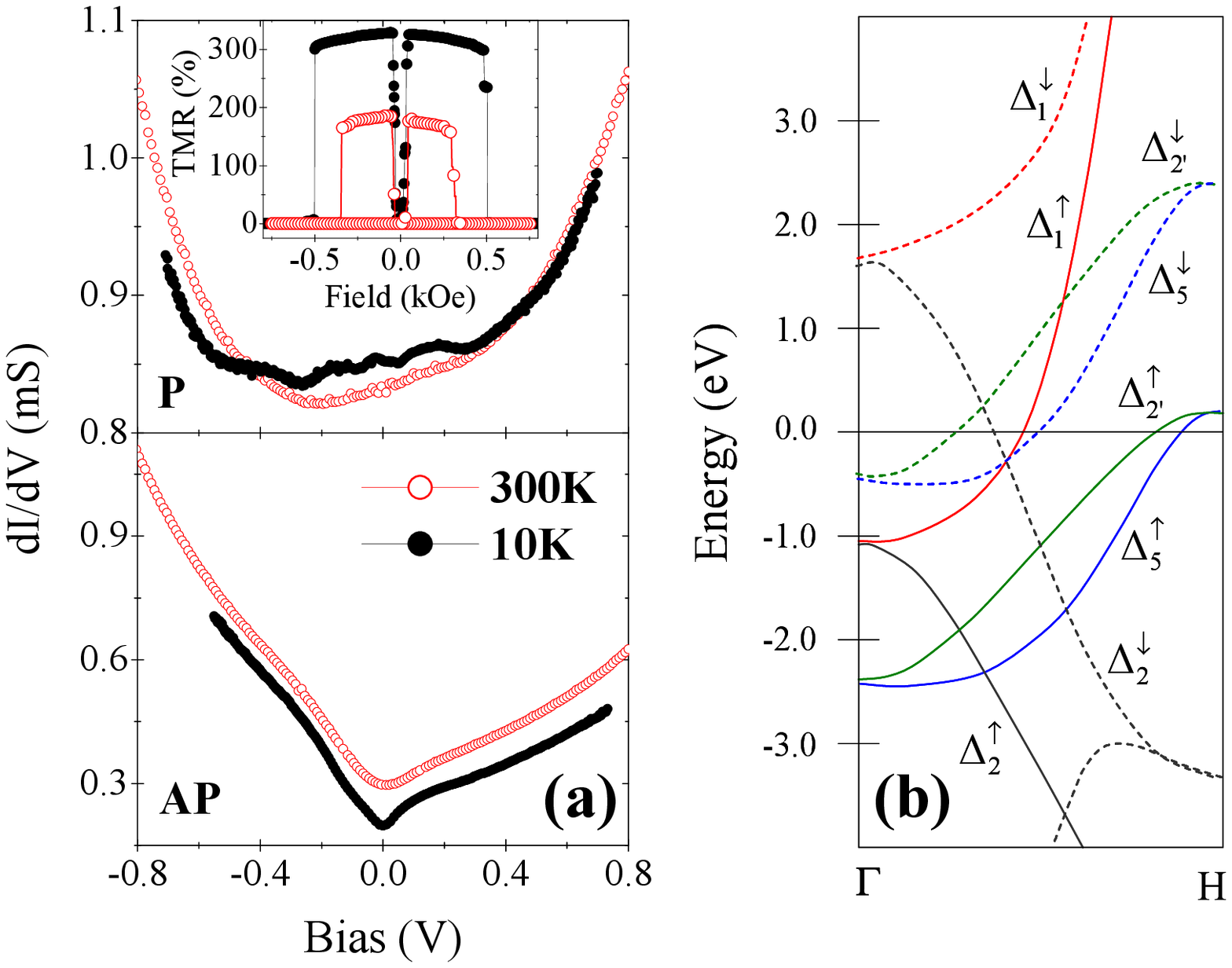}
\caption{}
\label{Fig2}
\end{center}
\end{figure}

\begin{figure}[htbp]
\begin{center}
\includegraphics[width=7.5cm]{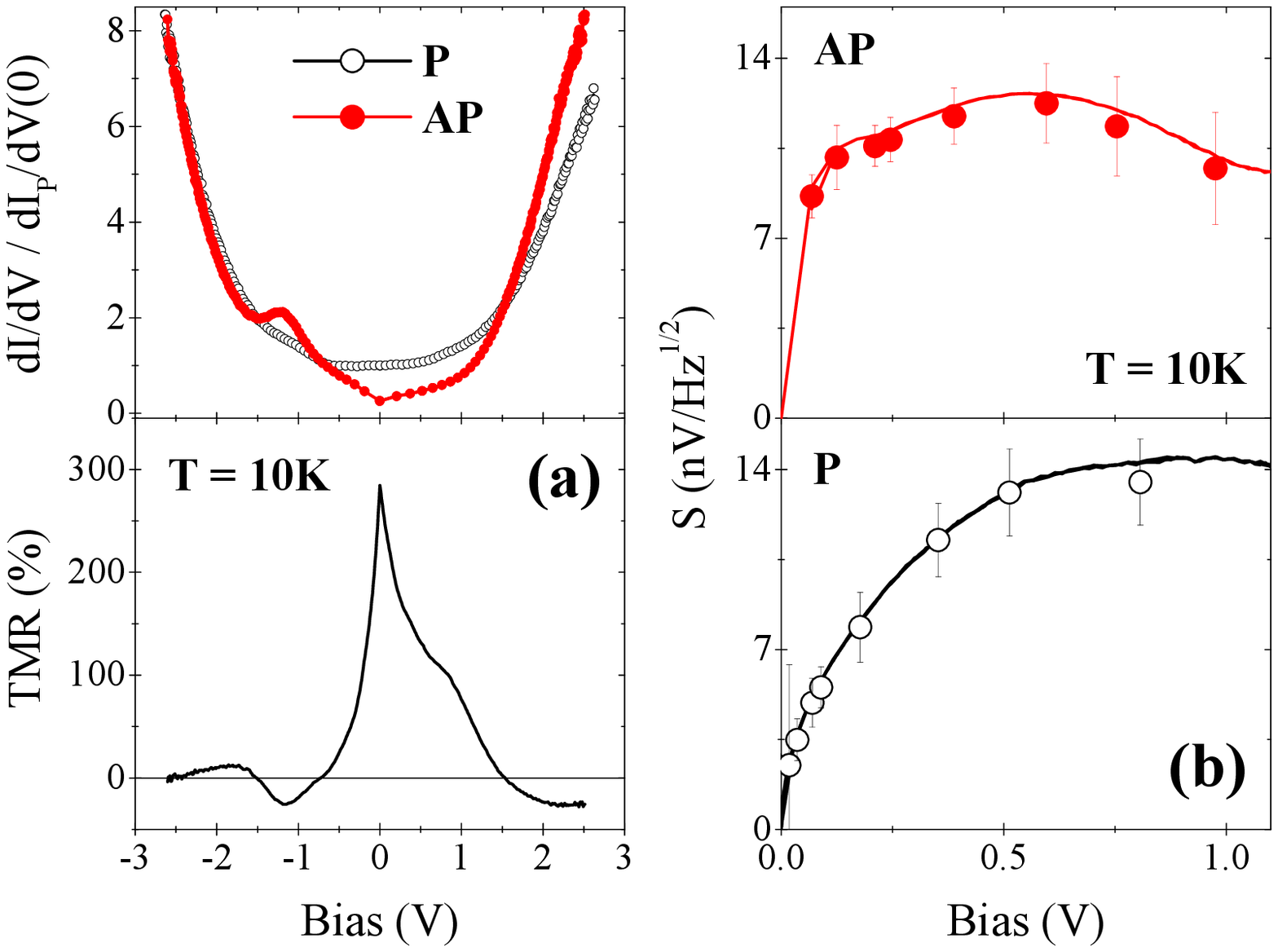}
\caption{}
\label{Fig3}
\end{center}
\end{figure}


\begin{thebibliography}{20}
\bibitem{IRS}
The interfacial resonance state of Fe has a $d_{z^2}$ orbital character and belongs to the $\Delta_1$ symmetry. However, due to the strong localization of the $d_{z^2}$ band, this state decays rapidly with the MgO thickness. Therefore, for large MgO thickness its contribution to the total AP conductivity is expected to be small.
\bibitem{width}
The spatial variation of the realistic electronic structure of the Fe-C/MgO interface (related to realistic morphology) will lead to a distribution in energy of the interfacial resonance.
\end{thebibliography}
\end{document}